\documentclass[pra,showpacs,showkeys,twocolumn]{revtex4}

\usepackage{amsmath}
\usepackage{amsfonts}            
\usepackage{amssymb}

\usepackage[colorlinks,urlcolor=blue,linkcolor=blue,pagecolor=blue,filecolor=blue,citecolor=blue]{hyperref}

\usepackage{mathrsfs}

\usepackage{eufrak}

\usepackage{verbatim} 

\usepackage{graphicx}
\usepackage{epstopdf}

\usepackage{bm}

\begin{document}
\title{Rotational structure of weakly bound molecular ions}

\author{Mikhail Lemeshko} 

\author{Bretislav Friedrich}

\affiliation{%
Fritz-Haber-Institut der Max-Planck-Gesellschaft, Faradayweg 4-6, D-14195 Berlin, Germany
}%

\date{\today}

\begin{abstract}
Relying on the quantization rule of Raab and Friedrich [Phys. Rev. A (2009) in press], we derive simple and accurate formulae for the number of rotational states supported by a weakly bound vibrational level of a diatomic molecular ion. We also provide analytic estimates of the rotational constants of any such levels up to threshold for dissociation and obtain a criterion for determining whether a given weakly bound vibrational level is rotationless. The results depend solely on the long-range part of the molecular potential. 
\end{abstract}


\maketitle

\section{Introduction}
\label{sec1}

Molecular systems bound by a potential which varies asymptotically with the inverse power of the distance, $r$, between their constituents
\begin{equation}
	\label{NoJpot}
	V(r) \overset{r \to \infty}{\sim}  -\frac{C_n}{r^n} \hspace{0.5cm} \text {with}~n>2 
\end{equation}
are amenable to an accurate analytic semiclassical (WKB) treatment, as long as the system's states are sufficiently ensconced within the potential energy well. However, for states near threshold for dissociation, the WKB approximation fails, as the system's classical action, proportional to momentum, no longer exceeds Planck's constant. And yet, it is the near-threshold states that have come to the fore recently, through the work in cold-atom physics where such states arise in photo- and magneto-association~\cite{RevFeshbach,RevFeshbachRudi,RevPhotoassoc} or other types of ``assisted'' collisions of (ultra)cold atoms~\cite{JensenHalo04} or atomic ions~\cite{GrierVuletic09, Idziaszek09, Zhang09, Jamieson09}. Therefore, a considerable effort has been expended at amending the WKB approximation to also allow for tackling near-threshold states analytically. A leading approach is that of H. Friedrich \emph{et al.}, who showed that, firstly, the bound-state eiegenenergies, $E_b$, can be expressed in terms of the quantization function, $F(E_b)$~\cite{FriedrichTrost2004}, which relates the state's integral quantum number $v$ to the generally non-intergral quantum number, $v_{th}$, of a state exactly at threshold, via
\begin{equation}
	\label{QuantFuncSimple}
	F(E_b) = v_{th} - v
\end{equation}
Secondly, they were able to find an explicit analytic form of the quantization function for attractive inverse-power potentials with $n=6$~\cite{HaraldPatrick08,PatrickHarald08}, and, most recently, with $n=4$~\cite{PatrickHarald09}. Note that the binding energy, $E_{b}=D-E_v$, with $D$ the dissociation energy and $E_v$ the energy of the vibrational level $v$, is thus positive, $E_{b}>0$. 

\begin{figure*}
\includegraphics[width=8.5cm]{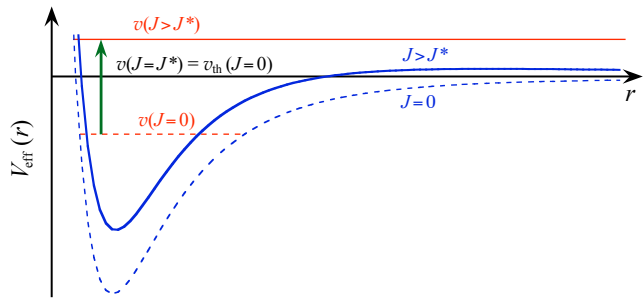}
\caption{(Color online) A schematic illustrating the role of the centrifugal term in the effective potential, $V_\text{eff}(r)  = -\frac{C_4}{r^4} + \frac{\hbar^2}{2 m} \frac{J(J+1)}{r^2}$. The energy splittings have been exaggerated for clarity. Shown is the position of a rotationless vibrational level, $v(J=0)$ (dashed line), as well as the position of the same level when pushed up by the centrifugal term to threshold, $v(J=J^\ast)=v_{th}(J=0)$ (full line, at threshold). When the rotational angular momentum $J$ exceeds the critical value $J^\ast$, the centrifugal term pushes the vibrational level above threshold, $v(J>J^\ast)$ (full line, above threshold), thus leading to dissociation.} \label{fig:eff_pot} 
\end{figure*}

In our previous work~\cite{LemFriRapid09}, we have shown that for each vibrational level, $v$, the rotational angular momentum, $J$, can take a critical value, $J^{\ast}$, such that the vibrational level is pushed up to threshold, thereby causing the level's binding energy to vanish. Hence the angular momentum $J$ in excess of $J^{\ast}$, $J>J^{\ast}$, dissociates the molecule, cf. Figure \ref{fig:eff_pot}. Furthermore, we have shown that the critical angular momentum is related to the quantization function by
\begin{equation}
	\label{Jast}
	J^\ast = F(E_{b}) (n - 2)
\end{equation}
The corollary of Eq. (\ref{Jast}) is that the number of rotational states supported by a weakly bound vibrational level of a dimer is given by the integer part of the critical angular momentum, $J_{max}=\text{Int}[J^\ast]$. By making use of the explicit form of the quantization function of refs. ~\cite{HaraldPatrick08,PatrickHarald08} for $n=6$, we were able to evaluate $J^\ast$ and estimate the rotational constant $B$ for near-threshold states of $^{85}\text{Rb}_2$. 

Here we take advantage of the recently derived explicit form of the quantization function for the $n=4$ long-range potential and analyze the rotational structure of highly-excited H$_2^+$ and $^{133}$Cs$_2^+$ molecular ions. The accuracy of the approach based on the quantization function is compared with that of an exact numerical solution of the corresponding Schr\"odinger equation.

\section{Rotational states of weakly bound molecular ions}
\label{sec2}
The quantization function for the $n=4$ case takes the form:
\begin{equation}
	\label{QuantFunction}
	F(E_{b}) = F_{th}(\kappa)+F_{ip}(\kappa) \Bigl[ F_{cr}(\kappa) + F_\text{WKB}(\kappa) \Bigr]
\end{equation}
where the individual terms, herein introduced for convenience, comprise a near-threshold dependence,
\begin{equation}
	\label{CoeffF1}
	F_{th}(\kappa) = \frac{2 b \kappa -(p \kappa)^2}{2 \pi \left[1 + (G_6 \kappa)^6 + (G_7 \kappa)^7 \right]};
\end{equation}
an ``interpolation'' term, which gives a smooth transition between low-$\kappa$ and high-$\kappa$ behavior, 
\begin{equation}
	\label{CoeffF2}
	F_{ip}(\kappa) =\frac{(G_6 \kappa)^6 + (G_7 \kappa)^7}{1+(G_6 \kappa)^6 + (G_7 \kappa)^7};
\end{equation}
a term which corrects the reflection phase due to the potential of Eq.~(\ref{NoJpot}),
\begin{equation}
	\label{CoeffF3}
	F_{cr}(\kappa) =-\frac{1}{4} +\frac{u_1}{2 \pi  \kappa^{1/2}}+\frac{u_3}{2 \pi  \kappa^{3/2}}+\frac{u_5}{2 \pi  \kappa^{5/2}}+\frac{u_7}{2 \pi  \kappa^{7/2}};
\end{equation}
and a pure WKB contribution,
\begin{equation}
	\label{CoeffF4}
	F_\text{WKB}(\kappa) =\frac{\kappa^{1/2}}{2 \sqrt{\pi}} \frac{\Gamma(\tfrac{3}{4})}{\Gamma(\tfrac{5}{4})}
\end{equation}
cf. Eq.~(38) of ref.~\cite{PatrickHarald09}. In Eqs.~(\ref{QuantFunction})--(\ref{CoeffF4}), the dimensionless wavenumber $\kappa$ is defined by 
 \begin{equation}
	\label{Varkappa}
	\kappa \equiv k \left(\frac{C_4 2 m}{\hbar^2} \right)^\frac{1}{2} = E_{b}^\frac{1}{2} C_4^\frac{1}{2} \frac{2 m}{\hbar^2}
\end{equation}
with $k = \sqrt{2 m E_{b}}/\hbar$ the wave vector and $m$ the diatomic's reduced mass. The parameters $b$, $p$ and $u$ in Eq.~(\ref{QuantFunction}) are listed in Table~\ref{table:Coefficients} (cf. Tables~I, II of ref.~\cite{PatrickHarald09}). Note that in order to avoid confusion with the rotational constant (defined below), we changed the symbols $B_{6,7}$, used in ref.~\cite{PatrickHarald09} for the adjustable-length parameters, to $G_{6,7}$. 

\begin{table}[t]
\caption{The $u_n$ coefficients of Eq.~(\ref{CoeffF3}).
The coefficients  of Eqs.~(\ref{CoeffF1}) and (\ref{CoeffF2}) are: $b \equiv 1$, $p^2 \equiv \frac{2\pi}{3}$, $G_6 =
1.622576$ and $G_7 = 1.338059$.}
\vspace{0.5cm}
\label{table:Coefficients}
\begin{tabular}{c c c c c c c }
 \hline
  \hline
\multicolumn{7}{ c  }{$n$ }   \\[3pt]
\hline
1 && 3  && 5 && 7 \\[3pt]
\hline\\[-10pt]
$\frac{5 \sqrt{\pi}~\Gamma(\tfrac{1}{4})}{48~\Gamma(\tfrac{3}{4})}$
&&  $- \frac{35 \sqrt{\pi}~\Gamma(\tfrac{3}{4})}{384~\Gamma(\tfrac{1}{4})}$
 &&  $\frac{475 \sqrt{\pi}~\Gamma(\tfrac{5}{4})}{3584~\Gamma(-\tfrac{1}{4})}$
 &&  $- \frac{63305
\sqrt{\pi}~\Gamma(\tfrac{7}{4})}{221184~\Gamma(-\tfrac{3}{4})}$\\[10pt]
  \hline
 \hline
\end{tabular}
\end{table}

The quantization function~(\ref{QuantFunction}) can be simplified by neglecting the $F_{th}(\kappa)$ and $F_{cr}(\kappa)$ terms and setting $F_{ip}(\kappa)$ to $\approx 1$, which results in the Leroy-Bernstein (LB) approximation~\cite{LeRoy70}. Neglecting only the $F_{th}(\kappa)$ and $F_{ip}(\kappa)$ terms yields the improved Leroy-Berstein approximation (iLB), which accounts for short-range deviations of the true potential from $V(r)$ of Eq.~(\ref{NoJpot}), see refs.~\cite{HaraldPatrick08,Jelassi0608}. The assumptions about the various terms of Eq.~(\ref{QuantFunction}) inherent to the approximations are listed in Table~\ref{table:Assumptions}.

Neglecting any coupling of the molecular rotation, we can estimate the rotational constant, $B$, from the rotational energy, $BJ^\ast(J^\ast+1)$, required to promote the vibrational level bound by $E_{b}$ to threshold
\begin{equation}
	\label{BrotConst}
	B = \frac{E_{b}}{J^\ast(J^\ast+1)}
\end{equation}
The values of the rotational constant $B$ obtained from Eq.~(\ref{BrotConst}) for $^{133}$Cs$^+_2$ and H$^+_2$ are listed in Tables~\ref{table:ResultsCs2ion} and \ref{table:ResultsH2ion} together with the essentially exact values, $B_{exact}$. The latter were calculated from
\begin{equation}
	\label{Bexact}
	B_{exact} = \langle v \vert \frac{\hbar^2}{2 m r^2} \vert v \rangle 
\end{equation}
with the vibrational wavefunctions obtained from a numerical solution of the Schr\"odinger equation for the potentials of refs.~\cite{Jamieson09} and \cite{Hilico00}.

\begin{table}[t]
\caption{Terms of the quantization function of Raab and Friedrich (RF), Eq.~(\ref{QuantFunction}), inherent to the LB and the iLB approximations. Also shown are the ranges of the reduced wavenumber $\kappa$ wherein the approximations apply. See text.}
\vspace{0.5cm}
\label{table:Assumptions}
\begin{tabular}{ c c c c c  c c c c}
 \hline
\hline
RF &&&& iLB  &&&& LB   \\[3pt]
\hline
All terms &&&& \begin{tabular}{c} $F_{th}=0$ \\  $F_{ip}=1$  \end{tabular} &&&& \begin{tabular}{c} $F_{th}=0$ \\  $F_{ip}=1$ \\  $F_{cr}=0$  \end{tabular}  \\[3pt]
All $\kappa$ &&&& $\kappa \approx 1$  &&&& $\kappa \gg 1$   \\[3pt]
  \hline
 \hline
\end{tabular}
\end{table}

The value of the binding energy $E_b$ for which a vibrational state is unable to support molecular rotation can be derived from Eq. (\ref {Jast}) with $J^\ast=1$. Hence the criterion for a level to be rotationless is 
\begin{equation}
	\label{CriterionEb}
	E_{b} < d_4 \frac{\hbar}{m^{1/2}~C_4^{1/4}}
\end{equation}
The parameter $d_4$ has the same value for all potential wells with an $1/r^4$ tail, namely  $d_4=2.9105$. Within the LB and iLB approximations,  the coeeficients $d_4^{iLB}=2.9096$ and $d_4^{LB}=0.7386$. These values come close to those given in Table III of ref.~\cite{LemFriRapid09}.

\section{Results and Discussion}

\subsection{ $^{133}$Cs$_2^+$ molecular ion}

As an example, we analyzed the rotational structure of the $^{133}$Cs$_2^+$ ion, for which an accurate potential energy curve was published recently~\cite{Jamieson09}.
Table~\ref{table:ResultsCs2ion} compares, for the last three vibrational levels of $^{133}$Cs$_2^+$, the values of the critical rotational angular momentum $J^\ast$ and rotational constant $B$, calculated from Eqs.~(\ref{Jast}) and~(\ref{BrotConst}), with exact results. The table also lists the values obtained by the Leroy-Bernstein and improved LeRoy-Bernstein approximations. One can see that for all the states considered, the predicted values of $J^\ast$ come close to the exact values, as do the values of $J^\ast_{iLB}$. However, $J^\ast_{LB}$, given by the purely semiclassical LB theory, are quite off the exact values. One can also see from Table~\ref{table:ResultsCs2ion} that the model's estimate of the rotational constants $B$ is within 25\% of the exact value. In the case of the LB approximation, the deviation of $J^\ast_{LB}$ from the exact value happens to be in the direction such that it improves the agreement of the LB rotational constant with the exact one; however, this improved agreement has to be regarded as serendipitous.
\begin{table}[t]
\caption{Critical angular momenta $J^\ast$ and rotational constants $B$ obtained for the last three vibrational states of the $^{133}$Cs$_2^+$ molecular ion in different approximations; $E_{b}$ and $B$ are given in $10^{-8}$ cm$^{-1}$. See also Table \ref{table:Assumptions} and text.}
\vspace{0.5cm}
\label{table:ResultsCs2ion}
\begin{tabular}{ c c c c c  c  c c  c   c c  c  c}
 \hline
\hline
$v$ && $E_{b}$ && $J^\ast$  & $J^\ast_{exact}$ & $J^\ast_{iLB}$ & $J^\ast_{LB}$ && $B$  & $B_{exact}$ & $B_{iLB}$ & $B_{LB}$  \\[3pt]
\hline
371 && 5.7   &&   0.63 &  0.63 & 0.63  & 1.01 &&  5.53  &  4.02  & 5.51 &  2.81 \\[3pt]
370 && 499.4  && 2.63  & 2.63  & 2.63 & 3.08 && 52.4   &  35.5  &  52.4  &  39.7 \\[3pt]
369 && 3731.1  && 4.62  & 4.64  & 4.62 & 5.10  && 143.5  &  96.4 & 143.5 & 120.0  \\[3pt]
  \hline
 \hline
\end{tabular}
\end{table}

\subsection{H$_2^+$ molecular ion}

We have also looked at the other end of the mass scale and tackled the rotational structure of the H$_2^+$ ion. The exact values of $J^\ast$ and $B$ were obtained by numerically solving the  the Schr\"odinger equation for the potential of ref.~\cite{Hilico00}. Table~\ref{table:ResultsH2ion} compares the exact critical angular momenta and rotational constants with those obtained from the models. Although the potential well of the hydrogen molecular ion is rather shallow, the near-threshold rotational structure is still governed by the long-range potential tail, as attested by a reasonable agreement of both $J^\ast$ and $B$ with the exact results. Since none of the last vibrational states lies in a ``pure near-threshold region," the LB and iLB approximations are not too far off either.

\begin{table}[t]
\caption{Critical angular momenta $J^\ast$ and rotational constants $B$ obtained for the three least-bound states of the H$_2^+$ molecular ion in different approximations; $E_{b}$ and $B$ are given in cm$^{-1}$. See also Table \ref{table:Assumptions} and text.}
\vspace{0.5cm}
\label{table:ResultsH2ion}
\begin{tabular}{ c c c c c  c  c c  c   c c  c  c}
 \hline
\hline
$v$ && $E_{b}$ && $J^\ast$  & $J^\ast_{exact}$ & $J^\ast_{iLB}$ & $J^\ast_{LB}$ && $B$  & $B_{exact}$ & $B_{iLB}$ & $B_{LB}$  \\[3pt]
\hline
19 && 0.707  &&  1.27 & 1.33 & 1.27 & 1.70 &&  0.24  & 0.15  & 0.24 & 0.15  \\[3pt]
18 &&  23.411 &&  3.60 & 3.70  & 3.60 & 4.07 && 1.41 & 0.92  & 1.41 & 1.14  \\[3pt]
17 && 153.67 &&  6.03 &  6.56 &  6.03 &  6.51 && 3.62  & 2.10  & 3.62 & 3.14  \\[3pt]
  \hline
 \hline
\end{tabular}
\end{table}

\section{Conclusions}

The advent of the physics of translationally cold atoms and molecules has secured a new prominence for long range interactions~\cite{Hutson09}. The $r^{-4}$ attraction features most notably in the ion-induced dipole interaction between an ion and and an atom or molecule, as well as in the  Casimir-Polder potential of an atom interacting with a surface~\cite{Babb05}. The bound states near threshold are of particular interest. This is because such states are non-classical, often spending most of their lifetime in the far-out, classically forbidden region of the potential~\cite{JensenHalo04}. Also, molecular species, whether ionic or neutral, in such states can be probed using nonresonant laser light, by ``shaking''~\cite{LemFriPRL09}. On the other hand, threshold states determine low-energy scattering behavior, subject to studies in traps~\cite{GrierVuletic09}.

Within this context (and beyond), it is helpful to possess the means the assess the rotational structure near threshold. Herein, we provide simple formulae that enable such an assessment to be carried out. These formulae yield the critical value of the rotational angular momentum needed for dissociation; the rotational constant; and a criterion for determining whether a given vibrational state can support molecular rotation.

\section*{Acknowledgments}

We are grateful to Harald Friedrich and Patrick Raab for discussions and to Gerard Meijer for encouragement and support.

\end{document}